\documentclass[a4paper,11pt]{article}
\usepackage{pos}

\usepackage{tikz}

\usepackage{url}
\usepackage{fancybox}
\usepackage[english]{babel}
\usepackage{times}
\usepackage{graphicx}
\usepackage{psfrag}
\usepackage{subfigure}
\usepackage{tabulary}
\usepackage{comment}

\usepackage{color}
\definecolor{darkred}{rgb}{0.4,0.0,0.0}
\definecolor{darkgreen}{rgb}{0.0,0.4,0.0}
\definecolor{darkblue}{rgb}{0.0,0.0,0.4}
\usepackage{braket}
\usepackage[T1]{fontenc}
\usepackage{amsmath}
\usepackage{amsthm}
\usepackage{url}
\usepackage[utf8]{inputenc}
\usepackage{dcolumn}
\usepackage{bm}
\usepackage{float}
\usepackage{url} 
\usepackage{grffile}
\usepackage{xcolor}
\usepackage{booktabs}
\usepackage{verbatim}

\usepackage{listings}
\newcommand{\be}{\begin{equation}}
\newcommand{\ee}{\end{equation}}
\newcommand{\beq}{\begin{equation}}
\newcommand{\eeq}{\end{equation}}
\newcommand{\bea}{\begin{eqnarray}}
\newcommand{\eea}{\end{eqnarray}}
\newcommand{\non}{\nonumber}
\newcommand{\bie}{\begin{small} \begin{itemize}}
\newcommand{\ie}{\item}
\newcommand{\eie}{\end{itemize} \end{small}}
\newcommand{\bbl}{\begin{block}{} \begin{center}}
\newcommand{\ebl}{\end{center} \end{block}}
\newcommand{\bref}{\begin{flushright} \begin{tiny}}
\newcommand{\eref}{\end{tiny} \end{flushright}}

\graphicspath{{./}}

\begin{document}

\title{%
Highly excited pure gauge SU(3) flux tubes
}
\ShortTitle{String spectrum}
\author*[a]{Pedro Bicudo}
\author[a]{Nuno Cardoso}
\author[a]{Alireza Sharifian}
\affiliation[a] {CeFEMA, Departamento de F\'{\i}sica, Instituto Superior T\'{e}cnico,
\\ Av. Rovisco Pais, 1049-001 Lisboa, Portugal
}
\emailAdd{bicudo@tecnico.ulisboa.pt}
\emailAdd{nuno.cardoso@tecnico.ulisboa.pt}
\emailAdd{alireza.sharifian@tecnico.ulisboa.pt}
\abstract{
Flux tube spectra are expected to have full towers of levels due to the quantization of the string vibrations. We study a spectrum of flux tubes with static quark and antiquark sources with pure gauge $SU(3)$ lattice QCD in 3+1 dimensions up to a significant number of excitations. To go high in the spectrum, we specialize in the most symmetric case $\Sigma_g^+$, use a large set of operators, solve the generalized eigenvalue and compare different lattice QCD gauge actions and anisotropies. 
}
\FullConference{%
The 38th International Symposium on Lattice Field Theory, LATTICE2021
\\ 26th-30th July, 2021
\\ Zoom/Gather@Massachusetts Institute of Technology
}
\maketitle

\section{Introduction - Motivation}\label{intro}

\begin{figure}
    \centering
    \begin{minipage}{0.45\textwidth}
        \centering
\includegraphics[width=1.\columnwidth,trim={0 30pt 0 210pt },clip]{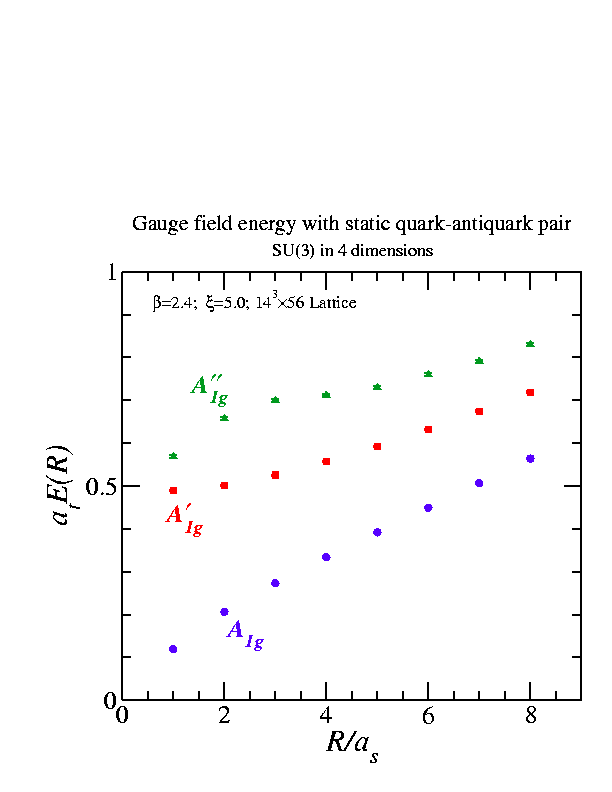}
\caption{\scriptsize Previous excited $\Sigma_g^+$ flux tube spectrum \cite{Morningstar:website}.
\label{fig:oldMorningstar}}    \end{minipage}\hfill
    \begin{minipage}{0.45\textwidth}
        \centering
\includegraphics[width=1.\columnwidth,trim={0 500pt 0  40pt},clip]{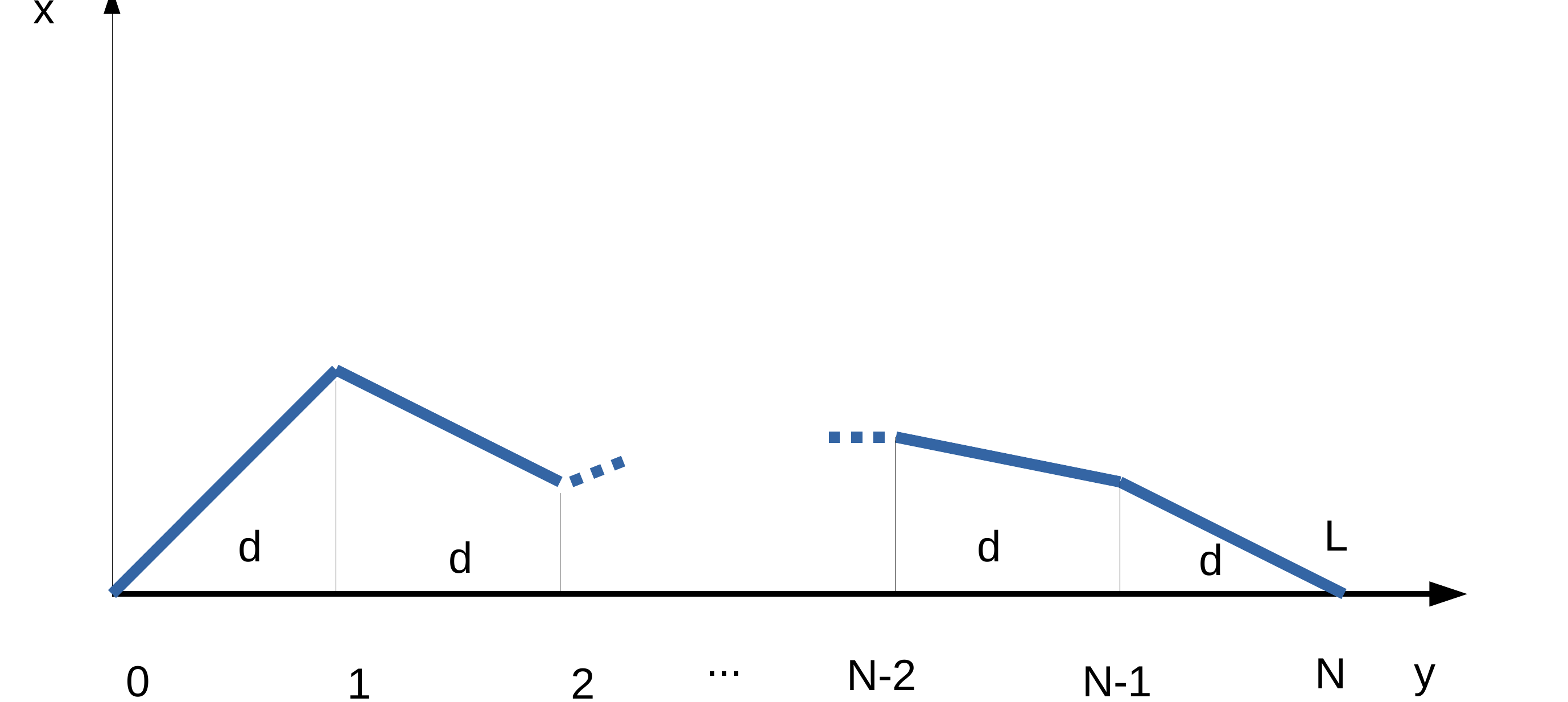}
\caption{\scriptsize The discretization of the string transverse quantum fluctuations is sufficient to compute the Coulomb term of the string spectrum.
\label{fig:discrete}}
    \end{minipage}
\end{figure}

So far only energy levels up to $N=2$ have been published for the pure gauge QCD flux tube \cite{Morningstar:website} as in Fig. \ref{fig:oldMorningstar}. Can we go higher in the spectrum? 

To understand why there is a spectrum, we notice that if we neglect the flux tube intrinsic width, it is equivalent to a quantum string. Flux tubes and strings have been studied for a long time.

\bie

\ie 
In 1911 Onnes discovered superconductivity, the Meissner effect was discovered in 1933.

\ie

In 1935, Rjabinin and Shubnikov experimentally discovered the Type-II superconductors. In 1950, Landau and Ginzburg, then continued by Abrikosov arrived at superconductor vortices, or flux tubes. 

\ie 
 When confinement was proposed for quarks inside hadrons in 1964 by Gell-Mann and Zweig, the analogy with flux tubes also led to a literature explosion in the quantum excitations of strings.

\ie 
The proposal of QCD in 1973 By Gross, Wilczek and Politzer shifted the interest back to particles.

\ie 
Nevertheless Lattice QCD  by  Wilson in 1974 was inspired in strings. 

\ie  The interest in strings returned in 1997 with Maldacena's and others  AdS/CFT correspondence.

\ie  The AdS/CFT and Holography has also been used as a model to compute spectra in hadronic physics.

\eie

\begin{figure}
    \centering
    \begin{minipage}{0.45\textwidth}
        \centering
\includegraphics[width=1.\columnwidth]{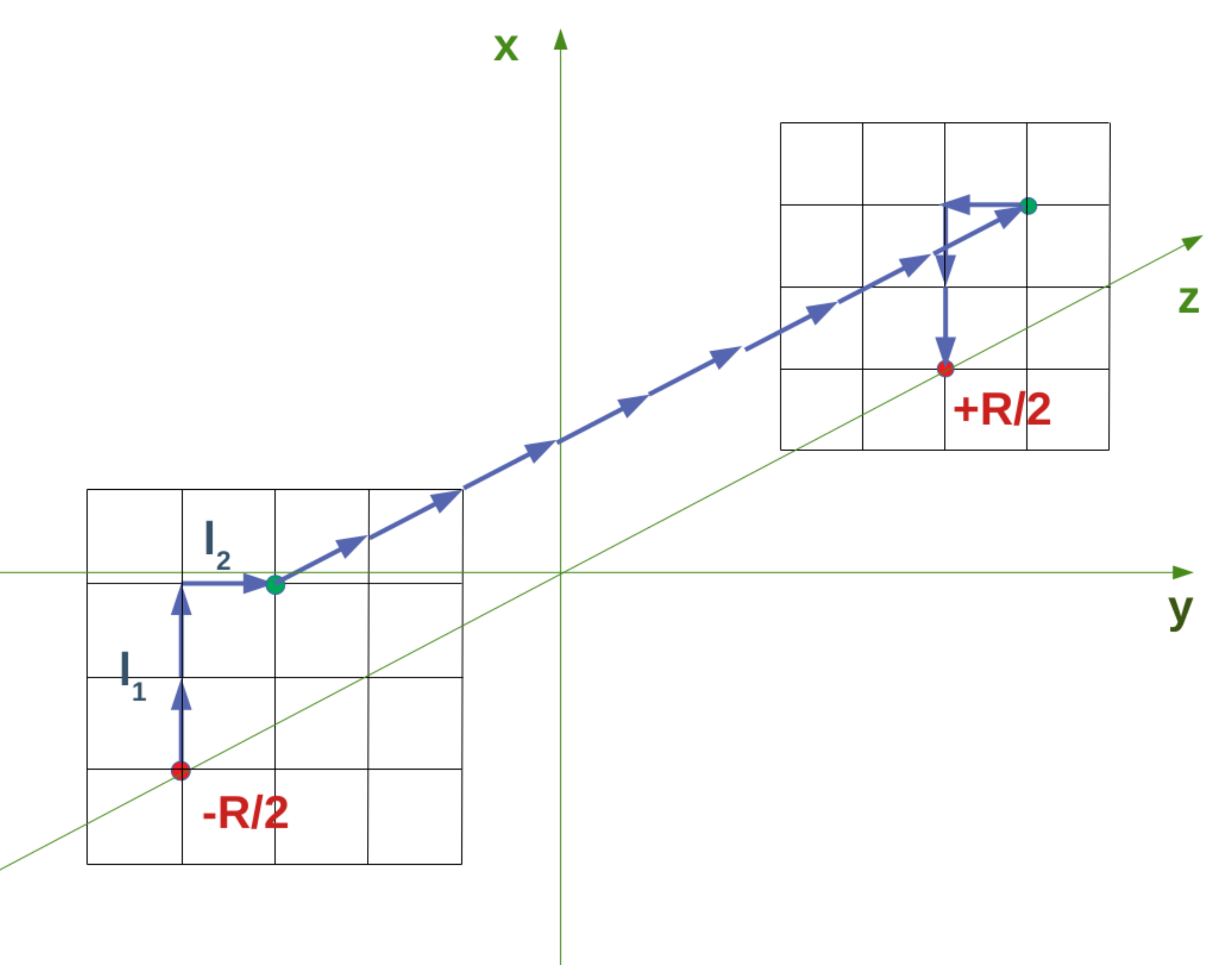}
\caption{\scriptsize Example of a lattice QCD spacial operator, in this case $O(2,1)$.
\label{fig:oper}}
   \end{minipage}\hfill
    \begin{minipage}{0.45\textwidth}
        \centering
\includegraphics[width=1.\columnwidth]{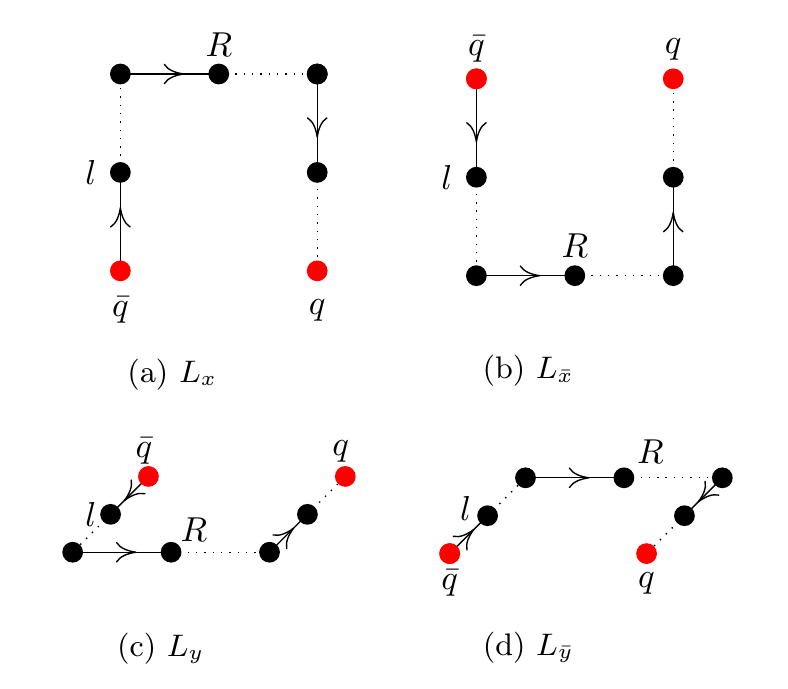}
\caption{\scriptsize The four sub-operators, spatial Wilson line paths from the antiquark to the quark, used to construct the gauge field operators $O(l,0)= {1 \over \sqrt 4} \left( L_x+L_{\bar x}+L_y+L_{\bar y} \right)$ at Euclidean time $t_0$. The inverse Wilson lines are used for the operators at time $t$.
\label{fig:MO0}}
    \end{minipage}
\end{figure}

An approximation to the flux tube spectrum is given by Effective String Theories (EST), say the Nambu-Goto model, which action is the area of the transverse bosonic  string surface in time and space, 
\be
S=-T \int d^2 \sigma \sqrt{- \text{det}\gamma} \ , 
\gamma_{\alpha \beta}={\partial X^\mu \over \partial \sigma^\alpha} 
{\partial X^\nu \over \partial \sigma^\beta}\eta_{\mu \nu}
\ee
it is classically equivalent to the Polyakov action which introduces an auxiliary  {\it einbein} field, 
$g_{\alpha \beta}$ to remove the square root,
\be
S=-{T \over 2}\int d^2 \sigma \sqrt{- \text{det} g} \, g^{\alpha \beta} {\partial X^\mu \over \partial \alpha} 
{\partial X^\nu \over \partial \beta}\eta_{\mu \nu} \ .
\ee
Its  spectrum for an open string with ends fixed at distance $R$ with Dirichlet boundary conditions is given by the Arvis potential  \cite{Arvis:1983fp},
\bea
V_{n}(R) &=& \sigma\sqrt{R^{2}+\frac{2\pi}{\sigma}\left(\sum_{a_i=1}^\infty n_{a_i}-\frac{D-2}{24}\right)} \ ,
\non  \\ \label{Arvis}
& =& \sigma R +\frac{\pi}{R}\left(\sum_{a_i=1}^\infty n_{a_i}-\frac{D-2}{24} \right)  +o \left( 1 \over R^3 \right)\ ,
\eea
where $a_i$ are the principal transverse modes, and the continuum field theory computation of the zero mode energy is obtained using the Riemann Zeta regularization \cite{Lovelace:1971fa,Brower:1972wj}, 
\be
1+2+3+\cdots = \sum_{n=1}^\infty  n \quad \quad \to \quad \quad \zeta(-1)= -{1\over 12} \ ,
\ee
which in lattice QCD is provided by the lattice regularization. The Coulomb approximation in Eq. (\ref{Arvis}), Lüscher term \cite{Luscher:1980ac}, can also be computed with a discretization of the String as in Fig.\ref{fig:discrete}.

However the Nambu-Goto model is certainly not the EST of QCD flux tubes.
\bie
\ie
There is a wider class of EST, the Nambu-Goto is just one of the possible EST \cite{Aharony:2009gg,Brandt:2021kvt,Bonati:2021vbc}.
\ie
Contrary to Nambu-Goto, the zero mode of the QCD flux tubes has no tachyon {\em with negative square masses} at small distances R.
\ie
There is lattice QCD evidence for an intrinsic width of the QCD flux tube \cite{Cardoso:2013lla}.
\ie
The QCD flux tube has a rich structure in chromoelectric and chromomagnetic field densities \cite{Bicudo:2018jbb,Mueller:2019mkh}.
\eie

Moreover, a quarkonium with flux tube excitations corresponds to  an exotic, hybrid excitation of a meson.
Thus we study the excited spectrum to learn more about the QCD flux tubes \cite{Bicudo:2021tsc}. We restrict to $\Sigma_g^+$ flux tubes, the most symmetric ones, to go as high as possible in the spectrum.

\section{Our lattice QCD framework for $\Sigma_g^+$ flux tubes}

\begin{table}[t!]
\begin{small}
\begin{tabular}{ccc|ccccccccc}
\hline
\hline
ens.  & action & operators & $\beta$ & Volume & $u_s$& $u_t$ & $\xi$ & $\xi_R$ & $a_s\sqrt{\sigma}$ & $a_t\sqrt{\sigma}$ & \# conf.\\
\hline
$O_1$	& Wilson & 11 $O(l_1,l_2)$ & 6.2 & $24^3\times 48$ & - & - & 1 & 1 & $0.1610$ &  $0.1610$ & 1180 \\
$O_2$	& 	Wilson & 11 $O(l_1,l_2)$ & 5.9 & $24^3\times 48$ & - &  - & 2  & 2.1737(4) & $0.3088(4)$ & $0.1421(2)$ & 2630 \\
\hline
$W_1$	& 	Wilson & 13 $O(l,0)$ & 6.2 & $24^3\times 48$ & - & - & 1 & 1 & $0.1610$ &  $0.1610$ & 2500 \\
$W_2$	&	Wilson & 13 $O(l,0)$ & 5.9 & $24^3\times 48$ & - &  - & 2  & 2.1737(4) & $0.3093(2)$ & $0.1423(1)$ & 2170 \\
$W_4$	&	Wilson & 13 $O(l,0)$ & 5.6 & $24^3\times 96$ & - & - & 4 & 4.5459(9) & $0.4986(4)$ & $0.1097(1)$ & 3475 \\
$S_4$	&	$S_{II}$ & 13 $O(l,0)$ & 4.0 & $24^3\times 96$ & 0.82006 & 1.0 & 4 & 3.6266(32) & $0.3043(3)$ & $0.0839(1)$ & 3575 \\
\hline
\hline
\end{tabular}
\end{small}
\caption{ Our ensembles, for the isotropic Wilson action and the improved anisotropic $S_{II}$ action. $\xi$ is the bare anisotropy in the Lagrangian and $\xi_R$ is the renormalized anisotropy. 
\label{tab:ensem}}
\end{table}

To compute the very excited spectrum, we use a large basis of spacial operators  composed of generalized Wilson lines, and example of their spacial part is illustrated in Fig. \ref{fig:oper} .

It turns out that using operators embedded only in axis planes, shown in Fig. \ref{fig:MO0}, we decrease the degeneracy of states in our spectrum.  This suppresses the $\Lambda_g$ states. Although we do not want to study them, due to the cubic symmetry of the lattice they may be as well generated by our operators.

The first step to compute the energy levels, is to diagonalise the Generalized Eigenvalue Problem for the correlation matrix of  our operators $O(l,0)$
\be
{\mathcal C}(t) v_n (t, t_0 ) = \lambda_n (t, t_0 ) {\mathcal C}(t_0 ) v_n (t_0) \ ,
\label{eq:GEVP}
\ee
for each time extent $t$ of the Wilson loop, and get a set of time dependent eigenvalues $\lambda_i(t)$.
With the time dependence, we study the effective mass plot
\begin{equation}
E_i \simeq  \log { \lambda_i(t) \over \lambda_i(t+1)} \ ,
\end{equation}
and search for clear plateaux consistent with a constant energy $E_i$ in intervals $ t \in [{t_i}_\text{ini}, {t_i}_\text{fin}]$ between the initial and final time of the plateau.

We use the anisotropic Wilson action \cite{Wilson:1974sk} computed with plaquettes, 
\begin{equation}
	S_\text{Wilson} = \beta\left(\frac{1}{\xi} \sum_{x,s>s'} W_{s,s'} + \xi \sum_{x,s} W_{s,t}  \right)
\end{equation}
where  $W_c = \sum_c {1\over3} \text{Re Tr}(1 - P_c )$, $P_{s,s'}$ denotes the spatial plaquette, $P_{s,t}$ the spatial-temporal plaquettes and $\xi$ is the (unrenormalized) anisotropy.
Moreover, to improve our signal we also resort to the improved anisotropic action $S_{II}$ developed in Ref. \cite{Morningstar:1996ze}, 
\bea
S_\text{II} &=& \beta \left( \frac{1}{\xi}     \sum_{x,s>s'}  \left[\frac{5 W_{s,s'}}{3 u_s^4} - \frac{W_{ss,s'}+W_{s's',s}}{12 u_s^6} \right]+   \right.
\non \\
& &\left. + \xi \sum_{x,s} \left[\frac{4 W_{s,t}}{3 u_s^2 u_t^2} - \frac{W_{ss,t}}{12 u_s^4 u_t^2} \right]   \right),
\eea
with $u_s = \langle {1 \over 3} \text{Re Tr} P_{ss} \rangle^{1/4}, \ u_t=1$.
$W_{ss,s'}$ and $W_{ss,t}$, instead of plaquettes, include $2\times 1$ rectangles.
The results with more excited states shown in the literature  
\cite{Morningstar:website}, have been obtained with this action.

The  anisotropy is used in order to have a smaller temporal lattice spacing $a_t$,  to obtain more precise plateaux for excited energies since we have more time slices for the same time intervals. 
 
For the $S_4$ ensemble and for the Wilson ensembles with anisotropy, we use MultiHit with 100 iterations in time followed by Stout smearing in space with $\alpha=0.15$ and 20 iterations.

We use GPUS, and we find it is more economical  to perform all our computations on the fly, rather than saving configurations. Our ensembles are summarized in Table \ref{tab:ensem}.

\section{Improved results for the $\Sigma_g^+$ spectrum}

\begin{figure}[t!]
\begin{centering}
\includegraphics[width=.45\columnwidth]{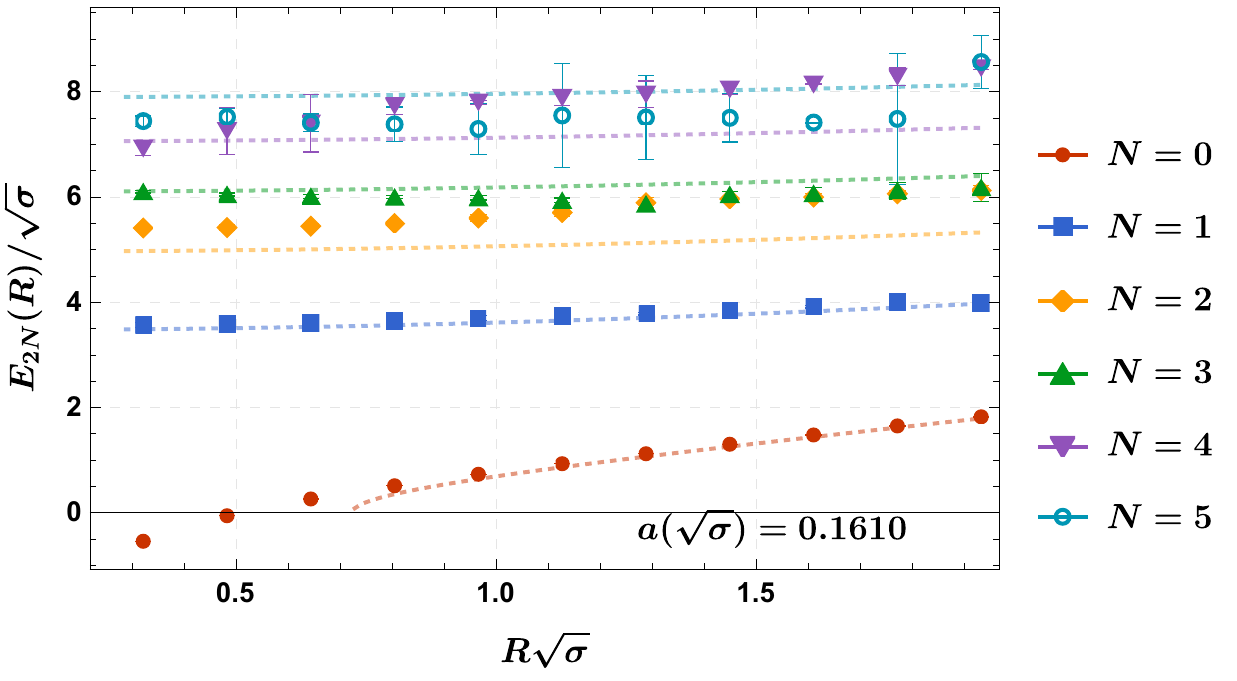} 
\quad
\includegraphics[width=.45\columnwidth]{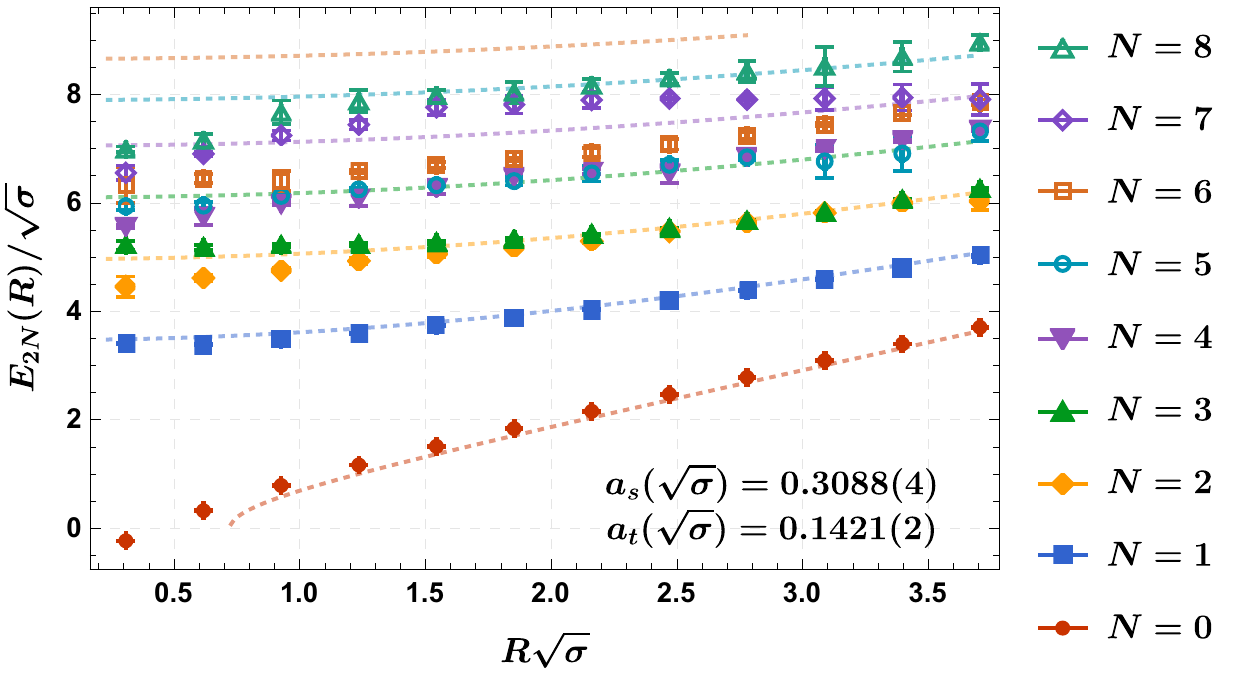}
\par\end{centering}
\caption{Results with the enlarged basis of 11 $O(l_1,l_2)$ operators with ensembles $O_1$ and $O_2$. The Nambu-Goto model spectrum is in dashed lines. }
\label{fig:pot_operaOl1l2}
\end{figure}

We obtain several plateaux with clear energy levels for each distance $R$. We first check for degeneracies.
From Nambu-Goto, we expect degeneracies:
\\
$n=0$, 1 state : $n_r=0$; 
\quad
$n=2$, 1 state : $n_r=1$;
\quad
$n=4$, 3 states: $n_r=2$ or l=4, 
\dots

Indeed, using off-axis operators, we find in Fig. \ref{fig:pot_operaOl1l2} nearly degenerate states $N=2$ and $N=3$, compatible with $n_r=2, \, l=0$ and $n_r=0, \, l=4$ states.

\begin{figure}[t!]
\begin{centering}
\includegraphics[width=.45\columnwidth]{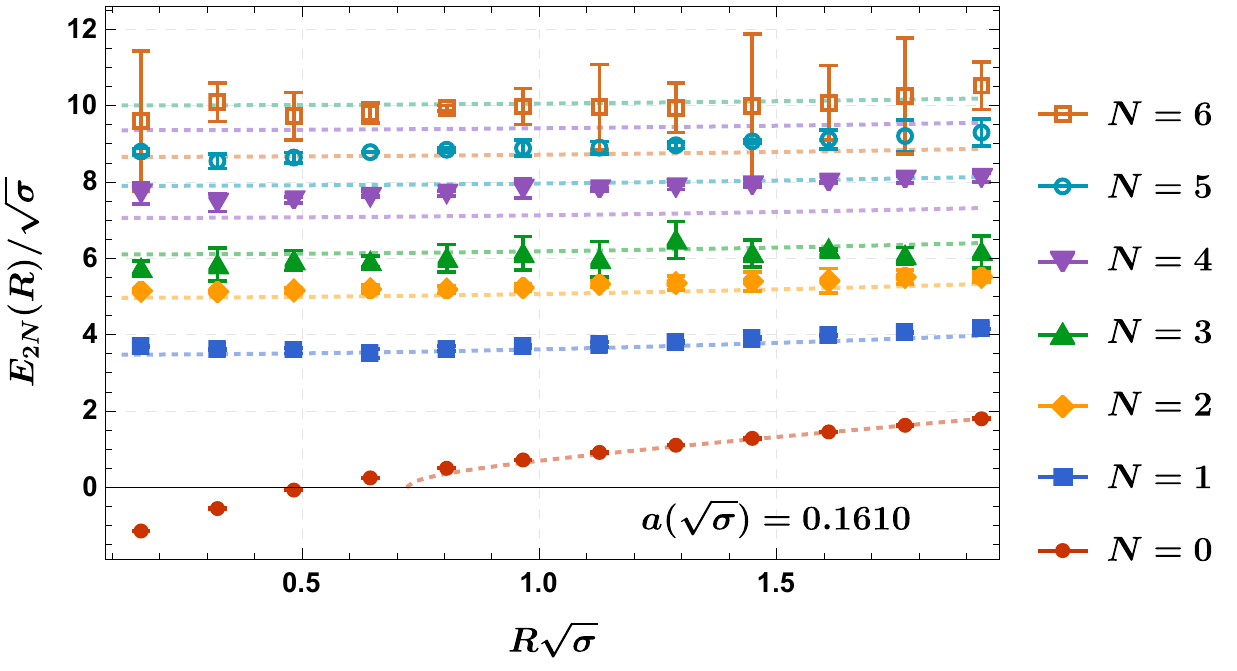}
\quad
\includegraphics[width=.45\columnwidth]{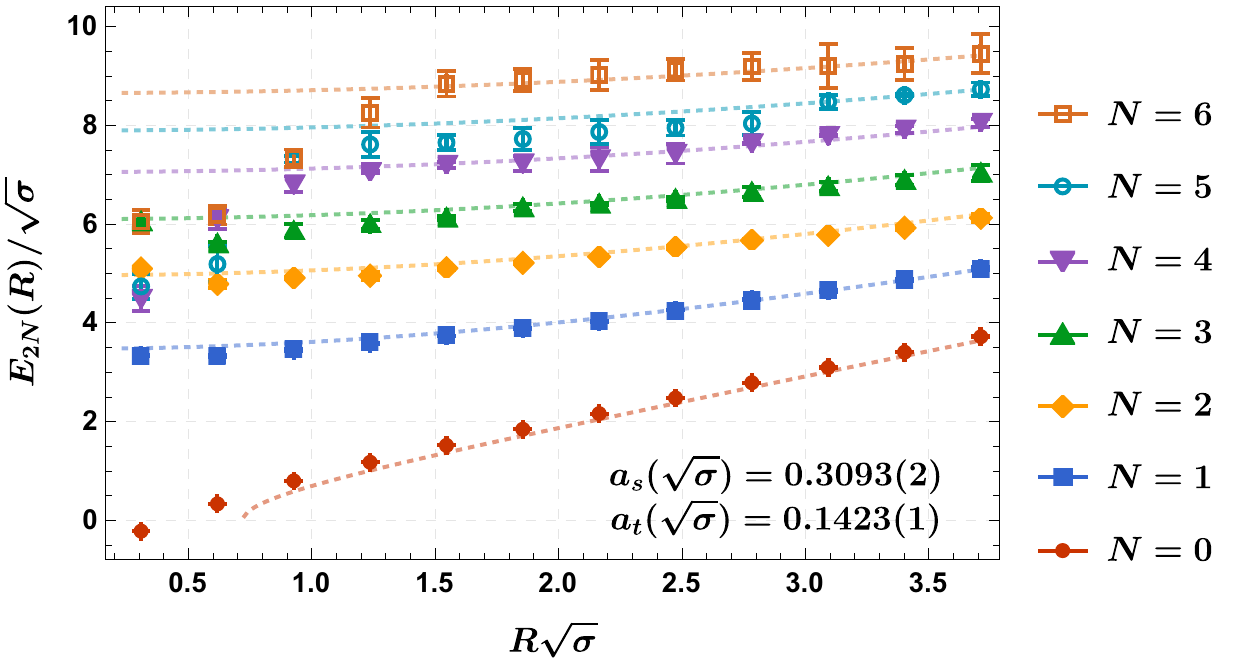}
\includegraphics[width=.45\columnwidth]{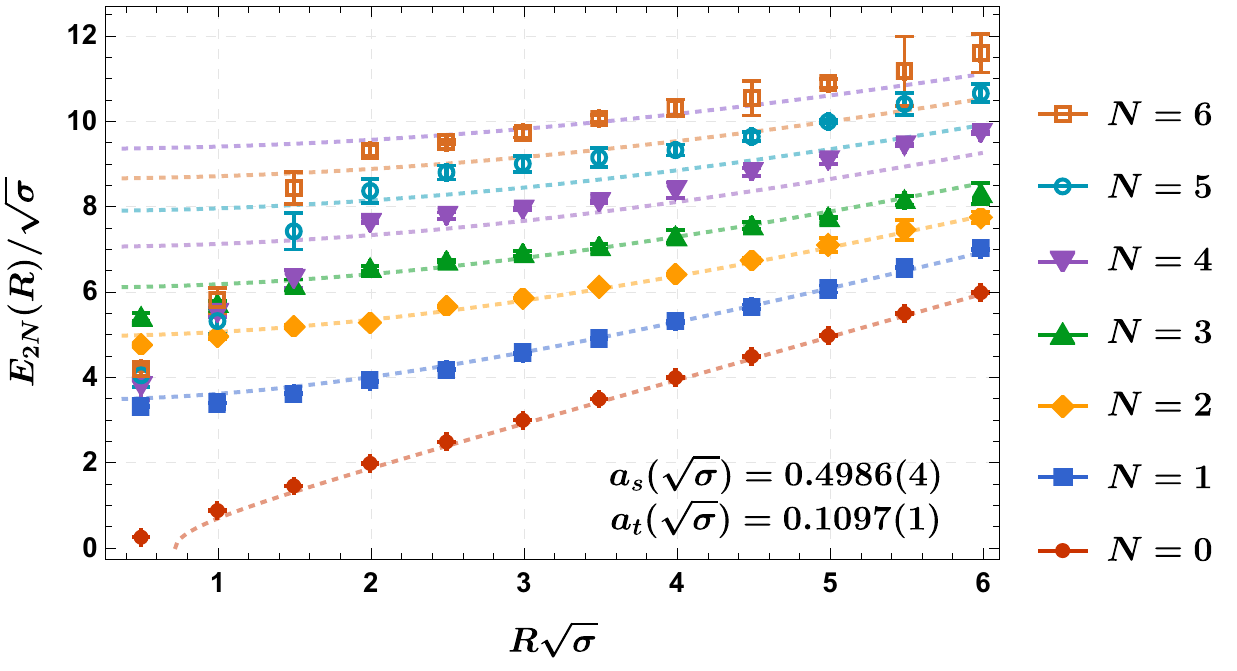}
\quad
\includegraphics[width=.45\columnwidth]{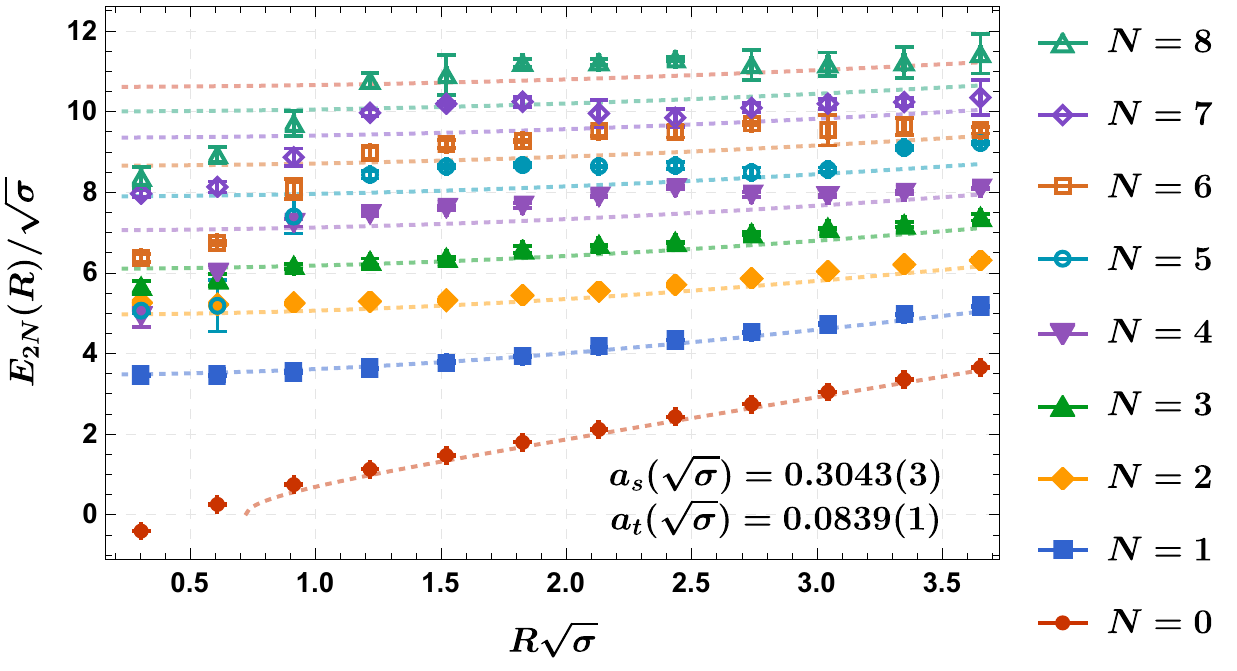}
\par\end{centering}
\caption{Results for the full spectrum, showing only the levels obtained with good Effective Mass Plots, from top to bottom respectively with ensembles $W_1$, $W_2$, $W_4$ and $S_4$. The Nambu-Goto model spectrum is in dashed lines.}
\label{fig:pot_opera_1244}
\end{figure}

Using only on axis operators, we are then able to suppress the $l=4$ degeneracy. At least for larger distances, we find up to   $N=8$ levels for the $S_4$ ensemble, shown in Fig. \ref{fig:pot_opera_1244}. 

At smaller distances, we are so far unable to avoid some degeneracy; possibly it is due to higher harmonics since $a_i=1, \, 3, \, 5 \cdots$ are as well $\Sigma_g^+$ states.

\section{Analysis of our $\Sigma_g^+$ spectrum}

\begin{figure}[t!]
    \centering
    \begin{minipage}{0.45\textwidth}
        \centering
\includegraphics[width=1.\columnwidth]{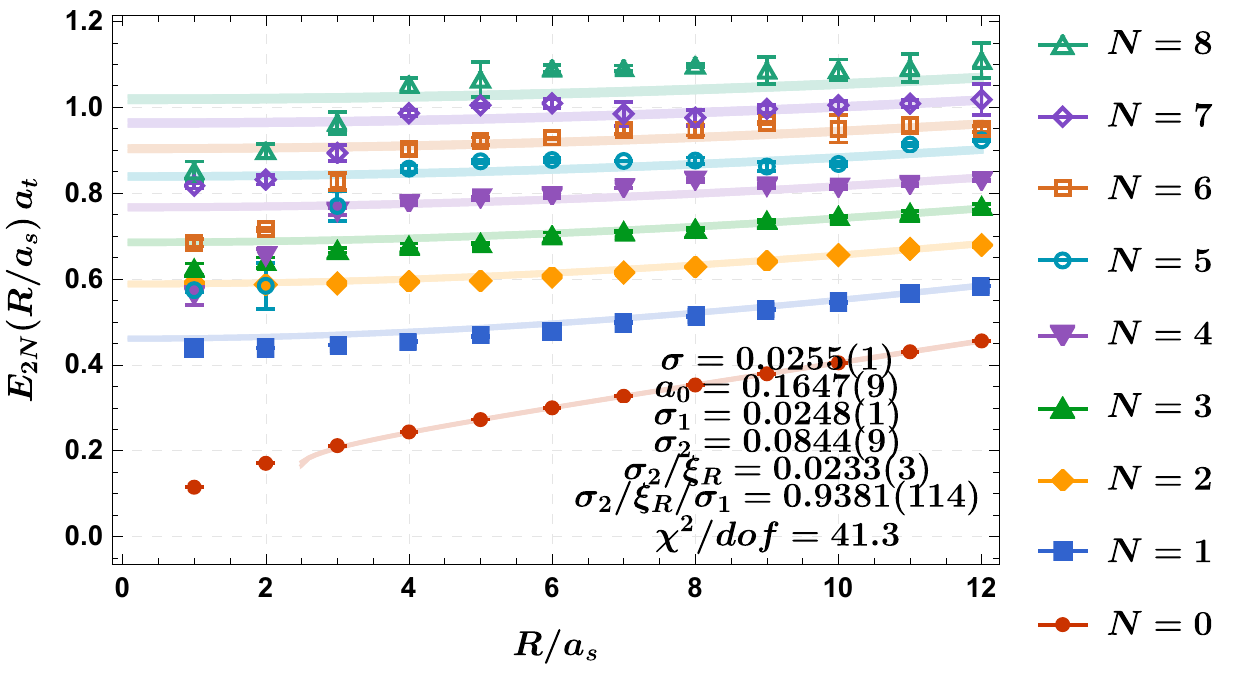}
\caption{\scriptsize Very excited spectrum of the QCD flux tube fitted by a generalized Nambu-Goto.
\label{fig:exci}}
\end{minipage}\hfill
    \begin{minipage}{0.45\textwidth}
        \centering
\includegraphics[width=1.\columnwidth,trim={0 5pt 0 0 },clip]{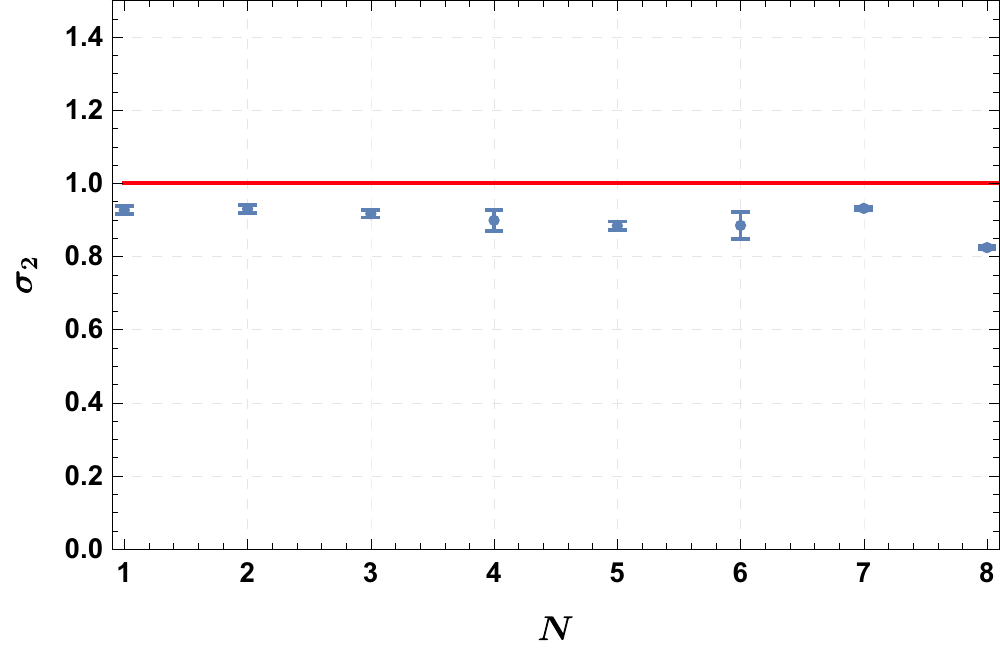}
\caption{\scriptsize Small deviation from the Nambu-Goto spectrum, $S_{II}$ action.
\label{fig:devi}}
    \end{minipage}
\end{figure}

We cannot fit accurately the very high spectrum with the Nambu-Goto spectrum which energy levels are more compressed than ours. We thus generalize the Nambu-Goto model using two different string tensions, the $\sigma_2$ replaces the $\sigma$ inside the square root.
With a global non-linear fit, shown in Fig. \ref{fig:exci}, we extract as well the renormalized anisotropy $\xi_R$ and the string tensions $\sigma, \, \sigma_2$.

Parametrizing the deviation to the Nambu-Goto spectrum, the second $\sigma_2$ is apparently slightly smaller than $\sigma$. We find a deviation, in  Fig. \ref{fig:devi}, of up to $10\%$ for the $S_4$ ensemble. A deviation is also present for the other ensembles albeit smaller In the $W_2$ ensemble.

\section{ Conclusion and discussion \label{sec:dis}}

We compute the potentials for several new excitations of the pure $SU(3)$ flux tubes produced by two static $3$ and $\bar 3$ sources, specializing in the radial excitations of the groundstate $\Sigma^+_g$.

Using a large basis of operators, employing the computational techniques with GPUs of Ref. \cite{Bicudo:2018jbb}  and utilizing different actions with smearing and anisotropy we go up to $N=8$ excitations.

In general the excited states of the the $\Sigma^+_g$ flux tubes are comparable to the Nambu-Goto EST with transverse modes, only depending on the string tension $\sigma$ and the radial quantum number $n_r$.

A detailed analysis shows a deviation of up to 10\% to the excited spectrum of the  Nambu-Goto Model. We leave the confirmation of this deviation for futures studies.

A subsequent feasible study is the computation of the widening in the different wavefunctions in the spectrum, and comparing them to from the zero mode widening. 

An important outlook would be the study of hybrid quarkonium resonances.

\vspace{15pt}
\noindent
{\bf Acknowledgements}
\vspace{15pt}
\\
We acknowledge discussions on flux tubes and constituent gluons with our colleagues Sergey Afonin, Gunnar Bali, Daniele Binosi, Bastian Brandt, Richard
Brower, Fabien Buisseret, Leonid Glozman, Alexei
Kaidalov, Felipe Llanes-Estrada, Vincent Mathieu, Colin
Morningstar, Lasse Müller, Fumiko Okiharu, Orlando
Oliveira, Emilio Ribeiro and Marc Wagner. We thank the
support of CeFEMA under the contract for R\&D Units,
strategic project No. 
UID/CTM/04540/2019, 
and the FCT project Grant No. 
CERN/FIS-COM/0029/2017.
NC is supported by FCT under the contract No.
SFRH/BPD/109443/2015
and AS is supported by the
contract No. 
SFRH/PD/BD/135189/2017.




\bibliographystyle{elsarticle-num}

\bibliography{excited}{}

\end{document}